\documentclass[12pt]{article}

\usepackage[top=4cm, bottom=4cm, left=3cm, right=2.5cm]{geometry}

\usepackage{setspace}
\usepackage{footmisc}

\makeatletter
\newcommand\iraggedright{%
\let\\\@centercr\@rightskip\@flushglue \rightskip\@rightskip
\leftskip\z@skip}
\makeatother
\usepackage{natbib,amsfonts,epsfig,amssymb,mathrsfs,times,authblk}
\usepackage{hyperref}
\usepackage[utf8]{inputenc}

\def\pn{\par\noindent}

\def\bs{\bigskip\pn}

\title{Validating the Universe in a Box}
\author[1]{Chris Smeenk}
\author[2]{Sarah C. Gallagher}
\affil[1]{Department of Philosophy and Rotman Institute of Philosophy, Western University}
\affil[2]{Department of Physics and Astronomy and Rotman Institute of Philosophy, Western University}

\begin{document}

\maketitle
\begin{abstract}Computer simulations of the formation and evolution of large scale structure in the universe are integral to the enterprise of modern cosmology.  Establishing the reliability of these simulations has been extremely challenging, primarily because of epistemic opacity. In this setting, robustness analysis defined by requiring converging outputs from a diverse ensemble of simulations is insufficient to determine simulation validity.  Instead, we propose an alternative path of structured code validation that applies eliminative reasoning to isolate and reduce possible sources of error, a potential path that is already being explored by some cosmologists.
\end{abstract}

\bs \textbf{Acknowledgments:} We have benefitted from questions and comments at PSA 2018, extended discussions with our co-symposiasts (in particular, Marie Gueguen), and constructive comments from two referees.  This paper was made possible in part through the support of grant \# 61048 from the John Templeton Foundation and the Natural Science and Engineering Research Council of Canada. The opinions expressed in this publication are those of the authors and do not necessarily reflect the views of the John Templeton Foundation.  

\section{Introduction}

Computer simulations have become an indispensable tool in modern cosmology.  While progress in the last century of physical cosmology can be attributed to theoretical innovations in concert with observational discoveries, in the next century progress on several questions will depend upon effective use of simulations.  Philosophers of science stand to learn from, and potentially contribute to, efforts to assess the validity of cosmological simulations as representations of the universe.%\footnote{For recent related work, see \citet{ruphy2011limits,Massimi2018,weisberg2018dark}.}     

Simulations are essential to understanding, inter alia, how galaxies of all types form and evolve. According to the $\Lambda$CDM model, formation of galaxies and large-scale structures begins with slight fluctuations away from a uniform background in the early universe.  Cold dark matter -- ``cold'' because the particles have negligible thermal velocities -- provides the scaffolding: initial ``seeds,'' consisting of local concentrations of dark matter (DM), grow through gravitational instability and merge to form larger DM halos.  These halos produce gravitational potential wells that attract baryonic matter, eventually leading to visible galaxies. This picture is widely accepted, but the devil is in the details.  Because it dominates the mass budget, DM plays a leading role, and comparisons of simulation outputs to observations may be the best way to discriminate among competing proposals regarding DM's nature.  Next-generation galaxy clustering and weak gravitational lensing surveys will yield higher precision measurements of large-scale structure, at a variety of scales.  Such surveys are being designed specifically to constrain cosmological parameters \citep[e.g.,][]{Euclid_science}, but measuring the true values with uncertainties of the cosmological parameters will depend on sophisticated statistical comparisons of large samples of galaxy properties to the expectations from a suite of simulations.  The success of these experiments thus fundamentally depends on the reliability of the simulations.

Current simulations have revealed several challenges in extrapolating $\Lambda$CDM from large scales, where it is most successful, to the scale of individual galaxies \citep[see also][]{Massimi2018}. When simulations yield results that conflict with observations, as they do in this case, should blame be attributed to an inadequate incorporation of all the necessary physics, or instead to artificial features introduced to render the models computationally tractable? Philosophers have discussed this challenge in other fields: how can we establish the reliability of simulations of complex systems, and to what extent do simulations promote physical understanding of systems that are otherwise difficult to observe directly?  State-of-the-art cosmological simulations are computationally intensive, and how a given feature in the models depends on the complex interplay of different assumptions (both physical and computational) is opaque.   How to justify simulation results and use them to guide further inquiry despite this opacity is a substantial challenge.  Our argument below focuses primarily on a critical assessment of one response, called ``convergence analysis'' or ``code comparison'' by cosmologists, and ``robustness analysis'' by philosophers.  There are close parallels in the arguments given by the advocates of this approach in both fields.  But more importantly, the limitations of this approach have been made particularly clear in the cosmology literature, and these cases help to identify two ways in which robustness analysis can fail.  Robustness may thus be considered necessary but not sufficient, and we discuss  ways forward for evaluating the success of cosmological simulations. 
%Need to reformulate here:  emphasis should be on how simulations guide inquiry overall, and the various roles they play

\section{The Universe \textit{In Silico}}

Astrophysicists are often simplistically separated into two camps: theorists and observers. However, the meaning of ``theorist'' in modern astrophysics is not the classical vision of an individual solving analytic equations, but rather typically involves developing and testing simulations of complex physical systems.  Cosmology as a field within astrophysics is no different, but the role of simulations is perhaps unusual.  Below are three specific ways in which the nature of the simulations is distinctive both inherently and in practice.  %(We focus here on simulations of large scale structure (size scales of Mpc to Gpc), rather than evolutionary simulations of individual galaxies or their immediate environs.)

\subsection{Seeing the Unseen}

Observationally, the distribution of matter in the early universe (circa 380,000 yrs after the Big Bang) is well constrained by the exquisite maps of the Cosmic Microwave Background Radiation (CMBR) made from data from {\em WMAP}, {\em Planck}, and other observatories.  %Though the bulk of the mass of the universe is in DM at all epochs and thus not directly observable, the baryons emitting light in the era of the CMBR were tightly coupled to the DM and thus excellent tracers of it.  
Though most of the mass of the universe at all epochs consists of DM, and is not directly observable, baryons emitting light in the era of the CMBR were tightly coupled to DM and are thus excellent tracers of it.  In the nearby universe, galaxy surveys that cover large swaths of the sky such as the Sloan Digital Sky Survey (\begin{small}\url{https://www.sdss.org}\end{small}) provide an in-depth view of the distribution of baryonic matter (traced by stars in galaxies).  Over cosmic time, baryons, unlike DM, can cool and condense by emitting light.  Therefore, baryonic matter grows much clumpier in its distribution in contrast to DM.  The difference in the physics dominating the two types of matter sets the stage for the role of simulations as a vehicle to gain access to information about the unseen DM.

Large scale structure simulations start with the initial conditions measured via the CMBR, and then propagate them forward in time by modeling gravity acting on DM within a $\Lambda$CDM cosmology.  Because DM dominates the mass, this suffices to predict the distribution of massive DM halos in the present epoch.  However, these halos are only observable with light-emitting baryonic tracers: galaxies.  Thus an additional step is required, which is to assign baryons to each DM halo, then evolve them forward in time with, e.g., semi-analytic prescriptions that attempt to describe the complicated baryonic physics such as that of star formation and supermassive black hole growth.  The databases\footnote{For example, \begin{small}\url{http://gavo.mpa-garching.mpg.de/Millennium}\end{small}.} holding the outputs for large scale structure simulations have two distinct types: catalogues of DM halos (from the first round of DM simulations), such as the Millenium Simulation \citep{millennium}, and catalogues of galaxies (from the subsequent incorporation of baryons into the models). The simulated galaxy catalogues are then compared to observations to test both layers of simulations \citep[e.g.,][]{Guo2011}.  The decoupling of the two steps allows multiple prescriptions of the baryonic physics to be investigated, as a mismatch between observed and simulated galaxy properties could arise from multiple sources.  Simulations thus tie surveys of observable galaxies to the distribution of DM halos that are otherwise invisible. 

\subsection{Watching the Movie}

From first formation to the present, galaxies change dramatically, as distant galaxies observed when the universe was a fraction of its present age are fundamentally different than the galaxies we observe locally.
% in properties such as size, shape, and amount of gas compared to stars.  
This evolution occurs over timescales of hundreds of millions to billions of years.  Astronomers cannot follow the transformation of, for example, an individual galaxy from a disordered collection of clumps to a well-organized spiral.
%Measurable time evolution in other physical systems modeled with simulations, by contrast, can occur on scales observable within a human lifetime.
Galaxy surveys capture only snapshots in time, and it is often not obvious which particular class of galaxies at a higher redshift includes the precursors to the types of galaxies such as spirals and ellipticals observed locally.  Simulation databases are therefore useful because they capture the evolutionary history of a particular galaxy or DM halo.  This allows a direct mapping between galaxies at any time slice of the simulation, and can be used to create evolutionary animations of what are otherwise static pictures of astrophysical objects or structures.  
 
 %\subsection{Looking for Relics}

\subsection{Guiding Observations}

Cosmologists also use simulations to motivate and optimize observing programs.
%A third characteristic mode in which cosmological simulations are used is to motivate and optimize observing programs.  
Though costly, supercomputer time is significantly less expensive than telescope resources.  %Telescope time is typically allocated via a competitive process decided by peer review panels, and research funding may accompany a successful observing proposal.  
Observatory time is both hard to get and valuable (for acquiring science data and enabling its analysis).  Simulations can serve two roles in determining allocation of telescope time.  First, predictions from simulations can motivate observing programs.  For example, simulations predicted swarms of low-mass dwarf galaxies around large galaxies.  The numbers predicted were not found around the Milky Way or other nearby massive galaxies -- the ``missing satellite problem'' -- but due to their low luminosity and surface brightness, these galaxies would be quite hard to detect without explicitly hunting for them.  Challenging observing programs were thus designed and carried out to target these elusive dwarf galaxies \citep[which are still ``missing''; e.g.,][]{SimonGeha2007}.  Second, for large observing programs, cosmological simulations are used to define the specifications, such as the area of sky and depth of images, that will enable significant constraints on the physics.  For example, the output from a simulation can provide the number of galaxies of a given type within a luminosity bin required to yield small enough uncertainties to constrain the parameter of interest.  Entire observatories such as the upcoming {\em Euclid} (\begin{small}\url{https://www.euclid-ec.org}\end{small}) space mission have survey programs designed in this manner.

\section{A Tale of Many Scales}

Cosmological simulations aim to describe structure formation on a range of scales, from large scale structure observed in the CMBR and large area galaxy surveys down to the colours and central mass profiles of individual galaxies.
%Accounting for the entire historical evolution of important structures, such as galaxies, is enormously complex. 
The relevant physics that must be incorporated into the simulation depends on the size scale of the phenomena under examination. Cosmologists often face a trade-off between resolution and scope. On the one hand, resolving individual galaxy halo substructure requires running high-spatial resolution simulations and including baryonic processes such as radiative cooling and star formation. To this end, a so-called ``zoom-in'' simulation must focus in detail on a single DM halo while the surrounding neighborhood remains at coarser resolution or is treated as a boundary condition.
%Consequently, as argued by Vogelsberger et al. (2015), zoom-in simulations do not offer a statistically relevant sample to compare with observations of large samples of galaxies.  Instead, they are intended to understand the complex properties of individual galaxies.
Zoom-in simulations are thus intended as in-depth case studies intended to reveal the physical origins of the structure and properties of individual galaxies such as our Milky Way.
On the other hand, large-scale simulations simulate a larger volume at lower resolution, but do not have the capacity to resolve the effects of the baryonic physics that impacts the central regions of DM halos most significantly. These lower-resolution simulations can be used to understand the observed large-scale mass distribution of DM halos, as probed, for example, by weak gravitational lensing surveys and galaxy clustering analyses.  Hence, defining precisely the rationale for the simulations --- such as supplementing sparse observations, understanding what the cold DM model predicts, unpacking how \textit{real} baryonic processes affect the evolution and inner structure of DM halos, or ruling out alternatives to $\Lambda$CDM --- is essential for determining which simulations should be used and under what circumstances simulations adequately represent the target systems they are modelling.

In this contribution, we focus on DM-only, large scale structure simulations.  At large scales, the goal is to describe the evolution of DM in a background cosmological model. To start, the early stages of the evolution of seed fluctuations can be described analytically as linear perturbations to an expanding universe model.  The basic dynamical equations for DM are the collisionless Boltzmann equation (for thermal velocities) and Poisson's equation (for the gravitational potential).  However, fluctuations enter the non-linear regime rapidly as density contrasts grow, and the gravitational dynamics can only be treated numerically, through $N$-body simulations.\footnote{$N$-body simulations describe gravity using Newtonian theory set in an expanding spacetime, which is assumed to be a good approximation to a relativistic treatment.}  Different ``gravity solvers'' provide schemes for numerically solving the dynamical equations.  $N$-body simulations treat the system as divided into representative ``particles'' (with some specified mass), compute the Newtonian forces acting on these particles, and then calculate their accelerations and velocities in discrete time steps.  The mass scale is set by available computational resources and the goals of the simulation, and the particles are (usually) theoretical constructs rather than representations of astrophysical bodies. 

The numerical problem is to solve the equations of motion, with the main computational cost associated with determining the gravitational potential for each particle.  Although initial simulations used a pair-wise calculation of the potential, this requires $O(N^2)$ steps and quickly becomes too computationally expensive with increasing $N$.  There are a variety of alternative, more efficient methods to find the potential. ``Tree'' methods calculate the gravitational potential by organizing the particle distribution in a hierarchical, tree-like structure, and then keeping the lowest order terms in a multipole expansion for the particles grouped together on leaves of the tree.  ``Particle mesh'' simulations, by contrast, introduce a three-dimensional mesh, calculate the density for each cell, and then determine the gravitational potential.  In both cases, a numerical integrator is needed to then calculate the positions and velocities of particles at some discrete time step.  Both approaches employ artificial components that do not directly map onto astrophysical objects or processes.  Tree methods use ``force-softening'', a prescription that renders the gravitational force weaker at small distances to avoid unphysically large accelerations from close encounters of the particles that would trigger numerical instabilities.  Particle mesh methods limit force resolution due to the mesh size, and particles interact through a mean field rather than through pointwise interactions.  These methods introduce new fundamental parameters:  the mass scale (set by the number of particles); the time step used in the numerical integrator; and a parameter fixing the force-softening (for tree methods) or a spatial scale (for particle mesh simulations).  

At smaller scales ($\le$ Mpc), other astrophysical processes have to be taken into account alongside gravity, including the radiative cooling and hydrodynamics of baryonic matter.  There is no natural separation of scales, as ``small-scale'' phenomena such as supernovae and supermassive black hole growth are expected to have a significant impact.  Such ``sub-grid'' processes are unresolved by the simulations, but have a profound impact on the structure at these small scales, such as the mass profiles in the inner regions of individual DM halos (the ``cusp-core'' problem) and the existence and number of the smallest DM halos (the missing satellites).  These are also the scales at the lower limits of numerical resolution.  
  
Altogether, these features conspire to make it difficult to build structure formation simulations spanning a large range of spatial scales and dynamical regimes, and furthermore to assess their overall reliability.  In particular, revealing the cause of mismatches with observations from simulation outputs is extremely challenging. Questions regarding the status of assumptions and input parameter values arise immediately:  at what mass and spatial scales, compared to these parameters, do these simulations give reliable descriptions of the gravitational dynamics of DM halos?  

\section{Through a Simulation, Darkly}

The sophistication and resolution of simulations has increased to the point that cosmologists often now present side-by-side comparisons of simulation outputs and observational data, challenging their audience to determine which is which.  Simulated wide-field maps from Illustris (\begin{small}\url{http://www.illustris-project.org}\end{small}), for example, are virtually indistinguishable from actual multi-colour images.  Yet determining the implications of such success is challenging due to the epistemic opacity of simulations. 

\citet{humphreys2009} argues that opacity is the philosophically novel feature of simulations.  In his sense, opacity reflects the inability of beings like us to survey or comprehend the full range of calculations performed.
%Reformulate:  not two sources of opacity, rather emphasis on what opacity amounts to.  
The way simulations are created and developed leads to a related type of opacity \citep[see, in particular][]{lenhard2019,winsberg2010}:  a lack of clarity regarding how to attribute praise or blame to the components of a simulation.  This generates the main obstacle to using simulations to gain physical understanding of target systems, and to fulfill the other roles of simulations in cosmology.

Ideally, a simulation code might be constructed from modules that each account for different physical effects relevant to the evolution of the system as a whole.  Within each module, the problem at hand must be transformed into an algorithm that can be feasibly implemented on a computer, by, for example, introducing a numerical solver in place of a set of differential equations.  Far from being dictated by theoretical considerations, there is scope for a variety of distinct implementations.  Simulations incorporate (following Lenhard) ``artificial components'' and ``kluges.'' The former are specific elements that lack theoretical motivation or even run directly counter to theory.  They are introduced to overcome computational limits, to counter problems like numerical instabilities, or to ensure good global behavior of the simulation.  Kluges are clumsy, \emph{ad hoc} -- and ubiquitous -- fixes required to make things work. (Kluge means an inelegant solution to a coding problem.) Including such elements may lead to new parameters that are not motivated physically, such as the length scale of a grid introduced in a discretized description of a continuous fluid.  The freedom to introduce artificial components and kluges is a feature, not a bug, as it enables the construction of working simulations. For any well-studied problem, there are generally competing simulations that employ different artificial components and kluges, with distinct parametrizations. This freedom also leads to opacity, however, because it obscures the contributions of different modules and other components of the simulation.      

Initial evaluation of simulations typically focuses on the global properties of their outputs, such as the distributions, masses, and colours of galaxies in a simulated universe.  ``Tuning'' the simulation to produce a particular kind of output -- such as the images cosmologists use to stump their colleagues -- exploits the parameterization freedom just described.  But it is also a further source of opacity that makes it more difficult to isolate and identify the contributions of the different components of the simulation.

Even if the modules include ``all the relevant physics,'' in practice they do not fit together seamlessly as one set of fully self-consistent mathematical equations. For example, modules within the code may not contribute as expected after the simulation has been tuned and modified with an eye toward global behavior, leading to what \citet{lenhardwinsberg2010} call ``fuzzy modularity.''  It is then unclear how one might generalize the initial success in fitting global proeprties, or use the simulation to learn more about other target systems in the computational domain. The initial success may be due to a delicate cancellation of errors that holds only for the chosen parameter values, for example.  

It is often difficult to tell the difference between such a cancellation of errors, and a case where the modules accurately represent key physics of the target system -- such as an accurate prescription for star formation. Only in the latter case would the simulation be expected to apply outside the domain for which it was explicitly tuned.  In the particular case of $N$-body cosmological simulations, determining whether the overabundance of low mass halos (the missing satellites) is a numerical artifact that might arise from force-softening or mass resolution issues \citep{vandenBosch2018}, or some aspect of DM or baryonic physics \citep{Bullock2017} that has not been properly included, has proven difficult.  The convergence of several models on this result is not sufficient justification for assuming they are correct; the possibility that there are dwarf galaxies in these numbers that are present but simply difficult to detect is shrinking. 

Acknowledging opacity need not lead to full-blown skepticism regarding simulations.  It should be possible to distinguish between artifacts of a simulation and properties that should be taken seriously.  Much as with \citet{franklin1989}'s defense of experiment, in our view scientists employ a variety of strategies for defending simulations, appropriate in different circumstances \citep{parker2008}.  These strategies can be most effectively employed when there are alternative sources of knowledge about the target system that can be used to ``benchmark'' simulations.  Two different sources of knowledge are particularly effective, yet frustratingly rare.  First, in some cases simulation outputs can be directly compared to analytic solutions of a fundamental set of equations, enabling a particularly clear assessment of the mathematical properties of the simulation.  Yet simulations are typically employed in exactly those cases where we lack such mathematical control. Furthermore, for simulations incorporating many different, coupled physical effects, identifying a core set of equations to which we should require fidelity can be difficult. Second, detailed experimental study of a target system can yield data not used in constructing the simulation; alternatively, experiments can be designed such that they explicitly aim to test key predictions of the models, such as looking for the ``missing'' dwarf galaxies.  Successes in accounting for data not used in the construction of the simulation support taking the initial success as not merely an artifact of flexibility and tuning.\footnote{We do not have space to argue further that predictions have more evidential value than accommodating the data by tuning the model. See \citet{Frisch2015} for a clear argument in favor of this claim, and references to related philosophical debates regarding climate simulations.}  Again, this kind of direct experimental validation may be unusual, because scientists most often employ simulations when relevant data are sparse. In cosmology, however, entire observatories are being designed with the aim of acquiring such validation. 

When empirical benchmarks are not readily available, background knowledge about the target system more typically consists of the results of other simulations.  For example, a new variant on a cosmological simulation would  include dark matter because of its previous successes as a key element in large-scale structure modeling. Validation of a simulation also includes comparing and contrasting it with an ensemble of other simulations. There are many choices to be made in constructing simulations that target the same systems, so it is natural to expect that the ensemble of existing simulations exhibits some degree of diversity.  In the context of cosmological simulations, the diversity should ideally be along distinct axes, including the computational methodology, the range of input tuning parameters, as well as the underlying physics in the cases where it is not well known.   

\section{Robustness and its Discontents}

Cosmologists and philosophers alike have suggested that convergence of an ensemble of simulations on a particular result supports taking it as a real feature of the target system. \citet{weisberg2006robustness} provides a recent influential defense of robustness analysis inspired by biological modeling practices and seminal earlier work by Levins and Wimsatt. The core idea is appealing: we should be more confident that a specific outcome, or feature of the target system, truly obtains if it is ``robust'' -- the common output of an ensemble of diverse simulations.   Even though any given model or simulation will contain idealizations and artificial components, agreement among independent simulations indicates that they share a common structure sufficient to generate the ``robust'' feature. To what extent can we rely on robustness analysis to draw the distinction between artifacts and real features of a target system, in light of the challenges discussed above?  The value of convergence of simulations has been the focus of active debate in cosmology.  Inspired by this work, we will briefly characterize two challenges to the sufficiency of robustness analysis and suggest an alternative approach.\footnote{Although we do not have space to engage with the philosophical literature on robustness in detail, the nature of these challenges is robust to variations among competing proposals.}  
%\footnote{On Weisberg's account, the analysis leads to a ``robust theorem'': \emph{ceteris paribus}, if the causal structure applies to a target system, then it will have the robust feature.}  

%... shift to convergence analysis, as practiced by cosmologists.   

%Need a general statement here doing two things:  (i) we don't need to engage in depth with the philosophical literature because these challenges are quite general...  

%identifies a common structure.     

The value of convergence clearly depends upon the nature of the simulation ensemble.  At one extreme, convergence over an \emph{ideal ensemble}, including all viable simulations for a given problem, would be particularly persuasive.  Such an ensemble of structure formation simulations would span differences in numerical implementation (such as particle mesh vs. tree codes), variations in uncertain physics (such as aspects of dark matter), artifical components, etc.  Convergence over such an ideal ensemble on a particular feature would establish that the details of implementation are irrelevant to the persistence of that feature, and support the idea that it follows from a ``common structure'' shared by all elements of the ensemble.  Yet it is clearly not the case that the currently available simulations span the space of possibilities in anything even remotely approaching this ideal.  So we must face the more challenging question, discussed in the case of climate simulations by \citet{parker2011}: how valuable is convergence for the ensembles we actually have?  What features of actual ensembles determine the evidential value of convergence?

%This is far too compressed, need to unpack here
There are (at least) two reasons that convergence over actual ensembles does not provide sufficient evidence to accept robust features, both stemming from the opacity emphasized above.\footnote{\citet{gueguen2019} considers two cases illustrating both failures, which we draw on here.}  First, cosmologists often consider ensembles spanning a relatively small region of parameter space.  Parameters appearing in $N$-body simulations, such as those setting the mass scale, time step, and force-softening (or spatial scale) are often chosen to ensure stability of the values of a set of outputs under variations in resolution.  Convergence studies establish the degree of sensitivity of simulation outputs to these parameter choices, and establish a range of ``safe'' values adopted in subsequent work.\footnote{Comparisons with observations enter into these considerations indirectly, in the choice of outputs for which stability under resolution variation is required; ``unsafe'' parameter values may still yield tractable simulations consistent with observations \citet{vandenBosch2018}. }  Yet this practice may generate the kind of cancellation of errors discussed above:  convergence may result from a delicate balance between numerical artifacts.  The difficulty of ruling this possibility out -- a consequence of the opacity of simulations -- undercuts claims that features are robust.
%Demanding convergence for a set of observables may obscure other aspects of the simulation.  
For an ensemble defined over a  broader range of parameter values, an apparently robust feature (such as large numbers of dwarf galaxy satellites) may disappear \citep[e.g.,][]{vandenBosch2018}.

Second, establishing that there is a robust feature is not sufficient to characterize the common structure producing it as really physical.  It is sometimes possible to identify precisely what aspects of an ensemble of simulations constitute the common structure responsible for a robust result, through detailed assessment of how the simulations respond to variations of input conditions and parameter values.
Benchmarking exercises as described above can be particularly powerful for identifying the common structure responsible for a robust result, and their utility underscores the limitation of robustness analysis on its own.
Furthermore, in some cases this common structure directly contradicts basic physical assumptions.\footnote{As \citet{gueguen2019} discusses, cosmologists showed that N-body simulations leading to a supposedly robust feature (``cuspy'' density profiles) violate conservation laws (for energy and other constants of the motion).} The simulations succeed due to --- rather than in spite of --- numerical artifacts.  Though the correct response in such cases may be to question the underlying physics, this possibility illustrates that the presence of a robust feature in itself does not preclude artificial components or kluges.

One way of refining robustness analysis in response to these failures would be to place further demands on the ensemble.  If simulations were fully modular, it would be possible to construct an ensemble by ``swapping out'' a particular part of the code with replacements that are sufficiently diverse.   Diversity is valuable in this case to the extent that it helps avoid the problems just noted.  If it might be the case that a particular artificial component of gravity solvers accounts for the robust feature, for example, then the ensemble should be enlarged to include a simulation that uses a different gravity solver.  An ensemble designed in this way would lead to a more compelling robustness argument.  An amendment to robustness analysis along these lines would then require an account of how to determine whether a given ensemble is sufficiently diverse, relative to a particular robust feature \citep[see also][]{schupbach2016robustness}.  This approach loses much of its utility and appeal, however, as modularity rarely holds.  In addition, actually carrying out the construction of simulations, in order to fill in this (potentially quite large) space of possibilities, strikes us as unfeasible, and would certainly be a substantial departure from current practice \citep{ruphy2011limits}.

The more compelling point is that we seem to have lost track of the reason why robustness is appealing.  To establish the reliability of simulations -- or any type of inquiry -- we need to identify possible sources of error, and then avoid them.  It is obviously unwise to rely on a single simulation, given our limited understanding of how its success in a particular domain can be generalized.  Robustness helps to counter such over-reliance.  But there are many other strategies that simulators have used to identify sources of error and rule them out \citep{parker2008}. First we must ask what are the different sources of error that could be relevant?  And what is the best case one can make to rule out competing accounts?

This eliminative approach can proceed, for example, by turning away from the cases for which the simulations have been tuned to match observations.  The contrast between numerics and physics is sometimes much sharper in contrived situations where many of the complexities of real-world systems have been removed.  Even if these do not correspond to a real target system, the simulation results may be subject to external validation -- such as through comparison to an analytic solution.  This comparison can render a clear verdict regarding whether a particular outcome results from numerical artifacts or the underlying physics.  This activity must be supplemented by a convincing argument that the differences between the simple case and complex target systems do not undermine extending the external validation to the cases of real interest.
%\footnote{In other words, we need assurance that what we have discovered in the contrived situation carries over to the real target systems.  This will be difficult to obtain, it may be more feasible than other responses to opacity.}   
Such arguments can effectively eliminate concern over a particular numerical artifact as opposed to the underlying physics, or vice versa.  This is a brief description of one strategy among many that cosmologists have employed to isolate numerical artifacts.  

\section{Conclusion}

Cosmology as a field places extreme demands on its simulations: they must enable tests of fundamental physics and support observational programs.  To date, structure formation simulations have successfully reproduced a wide range of features of the observed universe.  Because of the complexity and opacity of the simulations, however, the task of eliminating potential sources of error is extremely challenging.  Requiring that simulations converge on similar results does not provide adequate grounds for taking the simulations to have identified the causal structure of the target systems.  The approach we advocate differs from justifying simulations merely on the basis of their ability to capture some range of interesting observed phenomena.  A thorough program of eliminating the possibility of numerical artifacts is needed to justify using simulations to gain physical insight into the formation of galaxies.  Such efforts would be well justified, given the substantial resources being allocated to next generation observatories and the crucial role cosmological simulations play in both designing their specifications and enabling their science return.

%\bibliography{psa2018.bib}

\end{document}